\documentclass[11pt]{article}
\usepackage{latexsym}
\usepackage{amsmath}

\usepackage{graphicx}
\usepackage{subfigure}
\usepackage{amssymb}

\newcommand{\newc}{\newcommand}

\newc{\be}{\begin{equation}}
\newc{\ee}{\end{equation}}
\newc{\bea}{\begin{eqnarray}}
\newc{\eea}{\end{eqnarray}}
\newc{\beas}{\begin{eqnarray*}}
\newc{\eeas}{\end{eqnarray*}}

\newc{\pardt}{\partial_{t}}
\newc{\pardxi}{\partial_{i}}
\newc{\pardts}{\partial_{t^{*}}}
\newc{\pardxis}{\partial_{i^{*}}}
\newc{\pardxj}{\partial_{j}}
\newc{\pardxk}{\partial_{k}}
\newc{\pard}{\partial}
\newc{\ti }{\tilde}
\newc{\s }{\overline}

\newc{\sect}{\section}
\newc{\subs}{\subsection}

\newc{\defi}{\definition}
\newc{\prop}{\proposition}
\newc{\rem}{\remark}
\newc{\lem}{\lemma}
\newc{\exa}{\example}
\newc{\theo}{\theorem}
\newc{\coro}{\corollary}
\newc{\post}{\postulate}
\newc{\state}{\statement}

\begin{document}


\title{Non-equilibrium thermodynamics analysis of rotating counterflow superfluid turbulence}
\date{ }

\author{Michele Sciacca}

\maketitle
\begin{center} {\footnotesize Dipartimento di Metodi e Modelli Matematici Universit\`a di
Palermo, 
Palermo, Italy}




MSC 82D50; 76F99 
\end{center} \footnotetext{E-mail addresses: msciacca@unipa.it}

\begin{abstract}
In two previous papers two evolution equations for the vortex line
density $L$, proposed by Vinen, were generalized to rotating
superfluid turbulence and compared with each other. Here, the
already generalized alternative Vinen equation is extended to the
case in which counterflow and rotation are not collinear. Then,
the obtained equation is considered from the viewpoint of
non-equilibrium thermodynamics. According with this formalism, the
compatibility between this evolution equation for $L$ and that one
for the velocity of the superfluid component is studied. The
compatibility condition  requires the presence of a new term
dependent on the anisotropy of the tangle, which indicates how the
friction force depends on the rotation rate.
\end{abstract}

\section{Introduction}
Quantum turbulence is described as a chaotic motion of quantized
vortices in a disordered tangle, which in some occasions shows
diffusive or undulatory behavior \cite{VN}--\cite{D1}. The
measurements of vortex lines are described in terms of a
macroscopic average of the vortex line length per unit volume $L$
(briefly called {\it vortex line density} and which has dimensions
$length^{-2}$).

It is well-known in literature that when one apply a rotation to a
sample filled of helium II, a vortex array aligned to the rotation
axis is formed. In the steady state $L$ tends to $2 \Omega/\kappa$,
$\Omega=|\bf \Omega|$ being the modulus of the angular velocity and
$\kappa=h/m$ the quantum of vorticity ($m$ the mass of the $^4$He
atom and $h$ Planck's constant, $\kappa \simeq 9.97$
10$^{-4}$cm$^2$/s).

In the experiments on thermal counterflow, instead, an almost
isotropic tangle of quantized vortices is created, with
$L^{1/2}=\gamma V / \kappa$,  $\gamma$ being a dimensionless
coefficient which depends on temperature and $V$ the modulus of
the counterflow velocity defined below. These experiments are
performed by means of channel filled of helium II and heated at
one end. Using the two-fluid model, since only the normal fluid
component carries entropy and heat flow, its motion  will be from
the heated end to the opposite end. At the same time, to conserve
the mass the superfluid component must counterflow towards the
heater. Thus, a relative counterflow between the normal fluid and
superfluid components is established, characterized by $\bf V={\bf
V}_n-{\bf V}_s$  (${\bf V}_n$ and ${\bf V}_s$ being the velocities
of the normal and superfluid components averaged in a small volume
$\Lambda$). When counterflow velocity is higher than a critical
value then quantized vortices will be created.

Neglecting the influence of the walls, an evolution equation for $L$
under constant values of the counterflow velocity $\bf V$ was
formulated by Vinen \cite{V2}
\begin{equation}\label{Vin}%
\frac{dL}{dt}  = \alpha V L^{3/2}   - \beta \kappa
L^2,%
\end{equation}
 $\alpha$ and $\beta$ being dimensionless parameters. The steady state
solutions of equation \,(\ref{Vin}), apart from $L=0$, is $L=(\alpha
V/\beta\kappa)^2$.

There exists another version of (\ref{Vin}), the so-called
 alternative Vinen equation \cite{V3}--\cite{NF}
\begin{equation} \label{Vin-al} %
{dL\over dt} =A_1 {V^2\over\kappa} L-\beta
\kappa L^2,%
\end{equation}
which is also admissible on dimensional grounds. The steady state
solution is $L=A_1V^2/\beta \kappa^2$, in agreement with the
experimental results in completely developed turbulent regime.

The natural question which arises is what happens when the
combined situation of rotation and counterflow exists. Supported
also by experimental results \cite{SBD,YG}, this has addressed
many recent studies \cite{F1}--\cite{TBAM} in this direction, some
of which focusing the attention to extend Vinen's ideas to a wider
range of situations \cite{JM1}--\cite{Sciacca_PhisicaB_403}.

One of the open problems on this arguments regards the behaviour
of the turbulence when the rotation axis and the counterflow
direction are neither parallel nor orthogonal, two distinguished
cases already taken into account experimentally  \cite{SBD,YG}.
The parallel case, considered by Swanson {\it et al.}  in
\cite{SBD}, was already investigated in Ref.~\cite{JM1} --- where
the authors extended Vinen equation\,(\ref{Vin}) to this more
complicated situation --- and in
Ref.\,\cite{Sciacca_PhisicaB_403}, where instead the alternative
Vinen equation\,(\ref{Vin-al}) was extended. A comparison of the
theoretical studies and the experimental results was  only
admissible for the experiments performed by Swanson {\it et al.}.
A motivation for the analysis of the present paper could be the
behavior of a sphere filled of superfluid, which is submitted both
to a rotation around its own axis (such as neutron star) and to a
radial heat flux. Thus, the interaction between  rotation and heat
flow forming an arbitrary mutual  angle arises in a natural way in
these systems (see figures \ref{figura_stella}).

Because of the lack of experiments on rotating counterflow
situation with  $\bf V$  and $\bf \Omega$ forming  a generic angle,
some theoretical studies were made in Refs.\;\cite{JM-libro,JM2}
extending the modified Vinen equation, proposed in~\cite{JM1}, to
this case. There, a thermodynamic analysis to determine possible
coupling terms between the evolution equations of $L$ and $\bf V$
was performed.

In Ref.\,\cite{Sciacca_PhisicaB_403}, the extension of Vinen equation\;(\ref{Vin-al})
to the case of $\bf V$ parallel to $\bf \Omega$
was considered in order to explore whether these further arguments
are enough to establish which of both equations, (\ref{Vin}) or
(\ref{Vin-al}), is more suitable to describe actual experimental
results. Of course, to have a satisfactory version of the
macroscopic alternative Vinen equation (the same arguments are also
valid for the extension of Vinen equation), the respective
coefficients of all terms will have to be microscopically calculated
and found to coincide with macroscopic observations. These
perspectives are still far from the present abilities because of the
enormous difficulties arising to model a rotating tangle. Thus, a
combined effort in macroscopic and microscopic perspectives seems a
reasonable and promising way to proceed.

The main aims of this paper is to extend the modified alternative
Vinen equation, already proposed in\,\cite{Sciacca_PhisicaB_403},
to include  situations where the angle between $\bf V$ and $\bf
\Omega$ is not assigned (Section\,\ref{sez311}).  Furthermore, the
compatibility between the generalization of the alternative Vinen
equation  and the evolution equation for the velocity ${\bf V}_s$
is studied from the non-equilibrium thermodynamics  viewpoint
(Section\,\ref{Themodinamica}). The new equations
(equations\;(\ref{AltVinEqua_gen}) and (\ref{equapervstensor}))
model a class of physical phenomena where very few experiments are
available. However, the model appears to be confirmed by its
ability to describe related physical phenomena as shown in the
following sections. It is to note that the results of the present
paper need the support of the experiments to be confirmed and
improved. In particular, the model could be applied to the neutron
stars which are rotating compact self-gravitating objects whose
bulk consists of a superfluid neutron liquid at the huge density,
with a very small concentration of protons and electrons.

\section{Contribution of rotation to the evolution equation for the vortex line density: general case}\label{sez311}

In Ref.\;\cite{SBD}, Swanson {\it et al.}  and, in Ref.\;\cite{YG},
Yarmchuck and Glaberson  applied a heat flux to a rotating sample
showing a complex interaction between processes of formation and
destruction of vortices. Particularly, in the experiment performed
by Swanson {\it et al.} ($\bf\Omega$ and $\bf V$ collinear), they
observed that the effects of $V$ and $\Omega$ are not additive: in
fact, for sufficiently high values of $\Omega$, when the influence
of the walls can be neglected, the total vortex line density is
lower than $L_R+L_H$, $L_R$ and $L_H$ being the values of $L$ in
steady rotation and in steady counterflow superfluid turbulence,
respectively,  given by
\begin{equation}\label{LRLH}%
 L=L_R={2\Omega \over\kappa},\hskip0.5in L_H=
\gamma^2 \frac{V^2}{\kappa^2}, %
\end{equation} and the deviation
increases with $V$ and $\Omega$. Therefore, the rotation
facilitates the vortex formation, in the absence or for small
counterflow velocities, but it hinders their lengthening for high
values of $V$ and  $\Omega$. Furthermore, they observed that there
are two critical counterflow-rotation velocities $V_{c1}$ and
$V_{c2}$, proportional to $\Omega^{1/2}$ ($V_{c1}=
C_1\sqrt{\Omega}$, $V_{c2}=
 C_2\sqrt{\Omega}$, with $C_1=0.053$\,cm~sec$^{-1/2}$,
 $C_2=0.118$\,cm~sec$^{-1/2}$). For $V \le V_{c2}$ the length $L$ per unit volume of the vortex
lines is independent on $V$ and agrees with the
expression\,(\ref{LRLH}a), but with a slightly different
proportionality constant in the range $V_{c1} \le V \le V_{c2}$.
Finally, for $V \ge V_{c2}$, $L$ increases and becomes proportional
to $V^2$ at high values  of $V$.

The experiments performed, instead, by Yarmchuck and Glaberson in
\cite{YG} refer to an array of vortices, caused by a rotation
$\Omega$, which suffer the influence of an orthogonal counterflow.
They measure the gradients of temperatures and chemical potential
which confirms the presence of two critical velocities for the
counterflow velocity, the first corresponding to a depinning of
the vortices from the walls and the second one corresponding to
the transition to the turbulent state. The critical velocities are
proportional to $\sqrt{\Omega}$ as pointed out by Swanson {\it et
al.} in their experiments.

As mentioned in the Introduction, in the regime of high rotation
($0.2$ Hz $\le \Omega /2\pi \le 1.0$ Hz and $0\le V^2\le 0.2$
cm$^2$/s$^2$) and $\bf \Omega$ parallel to $\bf V$,
equation\;(\ref{Vin}) has been generalized into \cite{JM1}

\begin{equation}\label{VinEqua22} {dL\over dt}= -\beta  \kappa  L^2 +
\alpha_1\left[L^{1/2} - m_1 \frac{
\sqrt{\Omega}}{\sqrt{\kappa}}\right]V L + \beta_2\left[L^{1/2} -
m_2 \frac{ \sqrt{\Omega}}{\sqrt{\kappa}}\right]\sqrt{\kappa\Omega}
L,
\end{equation} where $m_1= {\beta_4}
/{\alpha_1}$ and $m_2= \beta_1/\beta_2$, and $\beta$, $\alpha_1$,
$\beta_1$, $\beta_2$ and $\beta_4$ are coefficients depending on
the polarization of the tangle, which was supposed to be a
function of $\Omega$ and $V$. As observed in\,\cite{JM1},
equation\,(\ref{VinEqua22}) describes  some of the most relevant
effects observed in the experiments of Ref.~\cite{SBD}.

Since  experimental results in pure counterflow are also described
by the alternative Vinen equation, in \cite{Sciacca_PhisicaB_403} a
generalization of it to rotating counterflow was proposed following
the main lines of Ref.~\cite{JM1}. This generalization was carried
out assuming that the counterflow velocity and the rotating axis
were parallel one to each other in order to compare theoretical
results with experiments  by Swanson {\it et al.}. This equation
assumed the following form
 \begin{equation}\label{AltVinEqua}%
 {dL\over dt} = -\beta  \kappa
L^2 +A_1 \left[L - \nu_1 {\Omega\over {\kappa}}\right]{V^2 \over
\kappa}
 +  B_1 \left[ L- \nu_2{\Omega\over {\kappa}}\right] {\Omega},%
  \end{equation}
and it is valid  only for $V$ collinear to  $\Omega$. In
Eq.\,(\ref{AltVinEqua}) coefficients $\beta$, $A_1$, $\nu_1$, $B_1$
and $\nu_2$ were supposed to depend on the polarization of the
tangle, \textrm{ i.e.} on $\Omega$ and $V$.


Now, following the general line of the paper \cite{JM2}
 equation\,(\ref{AltVinEqua}) is written to include the general
 situation in which vectors $\bf \Omega$ and $\bf V$ are not
 necessarily collinear. Looking at the terms of equation (\ref{AltVinEqua}), the
 only one which needs experimental examination is that proportional
 to $\Omega V^2$, because in the general case it could depend not
 only  on the absolute values of the two vectors $\bf \Omega$ and $\bf V$ but also
on the angle between them. A mathematically consistent version of
(\ref{AltVinEqua}) including both alternatives  is
\begin{eqnarray}\label{AltVinEqua_gen}
   {dL\over dt} &=& -\beta  \kappa
L^2 +L \left[\frac{A_1}{\kappa} {\bf V}\cdot {\bf U}\cdot {\bf
V}+B_1 {\bf \hat{\Omega}}\cdot {\bf U} \cdot {\bf \Omega}
\right] \nonumber \\
    && -\left[\frac{A_1V \nu_1}{\kappa^2}{\bf \Omega}\cdot \left(a_1 {\bf
\hat{V}}{\bf \hat{\Omega}}+a_2{\bf \hat{\Omega}}{\bf
\hat{V}}\right)\cdot {\bf V}+\frac{B_1\nu_2}{\kappa}{\bf
\Omega}\cdot{\bf U}\cdot {\bf \Omega}\right],
\end{eqnarray}
where ${\bf U}$ is the second order unit tensor, $a_1$ and $a_2$ are
two coefficients such that $a_1+a_2=1$, ${\bf \hat{V}}{\bf
\hat{\Omega}}$  and ${\bf \hat{\Omega}}{\bf \hat{V}}$ are diadic
products between ${\bf V}$ and ${\bf \Omega}$ such that the first
depends only on the absolute values of two vectors $V$, $\Omega$,
whereas the second one depends also on their angle, $\cos^2({\bf
V}{\bf \Omega})$.

As pointed out above, equation\,(\ref{AltVinEqua_gen}) includes
different situations and it becomes equation\,(\ref{AltVinEqua})
when one considers a rotating cylinder heated from the lower basis
or the higher basis.  A physically interesting situation arises when
two vectors are not parallel, for instance when one consider a
rotating cylinder heated from the wall, which means that $\bf V$ and
$\bf \Omega$ are perpendicular as in experiment performed in
Ref.\,\cite{YG} (these experiments will be discussed in the
Conclusions). In the general case, it is important to establish
 which term, $a_1$ and $a_2$, has to be included into
 equation\;(\ref{AltVinEqua_gen}). Since up to now we are not able
 to answer to this question, both coefficients $a_1$ and $a_2$ have to be considered.

The stationary solutions of equation\,(\ref{AltVinEqua_gen}) are
obtained requiring $d L/dt=0$, that is
\be\label{AltVinEqua_gen-sol}%
-\beta  \kappa L^2 +L \left(\frac{A_1}{\kappa} V^2+B_1 \Omega
\right) -\left[\frac{A_1V^2\Omega \nu_1}{\kappa^2}\left(a_1+a_2
\cos^2({\bf
\hat{\Omega}}{\bf \hat{V}})\right)+\frac{B_1\nu_2}{\kappa}\Omega^2\right]=0.%
\ee The case $\bf \Omega$ parallel to $\bf V$, that is the
stationary solutions of equation\,(\ref{AltVinEqua_gen-sol}) with
$\cos^2({\bf \hat{\Omega}}{\bf \hat{V}})=1$ and $a_1+a_2=1$ was
investigated in \cite{Sciacca_PhisicaB_403}. The main results of the
paper\;\cite{Sciacca_PhisicaB_403} are reported in order to extend
them to the general case dealt with in this paper.

\begin{figure}[h]
{\includegraphics[width=11cm]{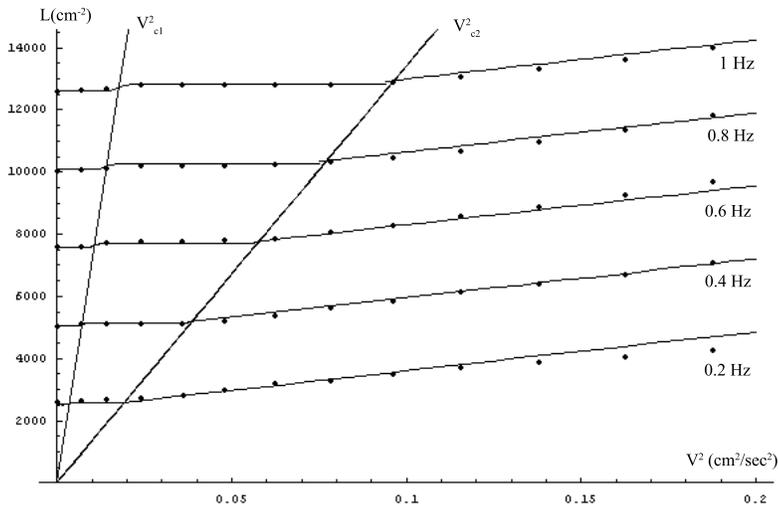}}
\caption{Comparison of the solutions (\ref{solstazL1AValue})
  (continuous line) with the experimental data by Swanson {\it et al.}
   \cite{SBD}. Figure from Ref.\,\cite{Sciacca_PhisicaB_403}.}\label{confronto}
\end{figure}

In the steady state ($L$ and $\bf V$ constant), under the
constraint
\be\label{condizioni_coeff}%
\nu_2=\nu_1-\nu_1^2\frac{\beta}{B_1},%
\ee  the stationary solutions of equation (\ref{AltVinEqua}) can
be written
\begin{equation}\label{solstazL1}%
L=L_1^\parallel =\nu_1{\Omega\over\kappa}, %
\end{equation}
\begin{equation}\label{solstazL2} %
L=L_2^\parallel= {A_1\over\beta}
{V^2\over\kappa^2}+\left({B_1\over \beta}
-\nu_1\right){\Omega\over\kappa}.%
 \end{equation}
Solutions  (\ref{solstazL1}) and (\ref{solstazL2}) are two families
of straight lines in the plane $(V^2,L)$, as plotted in
Fig.~\ref{confronto}: the first of them (equation (\ref{solstazL1}))
parallel to the $V^2$ axis and the second one (equation
(\ref{solstazL2})) with a slope independent of $\Omega$. Applying
the stability analysis to these solutions,
solution\,(\ref{solstazL1}) is stable if $V$ is lower than
\begin{equation}\label{VeloCrit2}%
V_{c2}^2= {\beta\over A_1}
{\nu_1^2-2\nu_1\nu_2 \over \nu_1-\nu_2} {\Omega \kappa},%
\end{equation}
while, for $V$ higher than $V_{c2}$  the solution\,(\ref{solstazL2})
is stable. The obtained value of the critical
velocity\,(\ref{VeloCrit2}) corresponds to the  interception of the
two  straight lines (\ref{solstazL1}) and (\ref{solstazL2}) in
Fig.\ref{confronto}. It also corresponds to the second critical
counterflow-rotation velocity observed in the experiments of
Ref.~\cite{SBD}. As we see, this critical velocity scales as
$\sqrt{\Omega}$, in agreement with experimental observations.

To describe the existence of the first critical velocity $V_{c1}$ in
the experiments, and also the small step in $L$, in
Ref.\,\cite{Sciacca_PhisicaB_403} $\nu_1$ was assumed dependent on
$\Omega$ and $V$ as
\be\label{espreni1}%
\nu_1= A\left\{1-B \tanh \left[ N'\left({
{k\Omega} \over V^2} -C\right)\right] \right\}.%
\ee There,  $A=2.018$, $B=0.0089$ and $C=0.355$ were found in such
a way that for $V^2 \ll V_{c1}^2  =  {1\over C}{k\Omega}$, it
results $\nu_1\simeq A-BA=2$ and for $V^2\gg V_{c1}^2$,
$\nu_1=A+BA=\nu^{max}=2.036$, while the constant $C$ is related to
 the critical value  $V_{c1}$  of  the counterflow velocity by $ V_{c1}^{2}={1\over C}
{\kappa\Omega},$ whose experimental value is $V_{c1} =0.053
\sqrt\Omega$ cm~sec$^{-1/2}$. In (\ref{espreni1}), $N'$ is a
phenomenological coefficient characterizing the rate of growth of
$L$  near $V_{c1}$ and for it the value $N'=22$ proposed by
Tsubota {\it et al.} in Ref.~\cite{TBAM} was assumed.

Now, let us investigate on the stationary solutions of
equation\;(\ref{AltVinEqua_gen}). Thus, assuming that
relation\;(\ref{condizioni_coeff}) is still valid when one
substitutes $\nu_1$ with $\ti \nu_1=\nu_1 (a_1+a_2 \cos^2({\bf
\hat{\Omega}}{\bf \hat{V}}))$, that is
\be\label{condizioni_coeff-gen}%
\nu_2=\ti \nu_1-\ti\nu_1^2\frac{\beta}{B_1},%
\ee then (\ref{AltVinEqua_gen-sol}) has the following  stationary
solutions
\begin{equation}\label{solstazL1-gen}%
L=L_1=\ti\nu_1{\Omega\over\kappa}=(a_1+a_2 \cos^2({\bf
\hat{\Omega}}{\bf \hat{V}})){\Omega\over\kappa}, %
\end{equation}
\begin{equation}\label{solstazL2-gen} %
L=L_2= {A_1\over\beta} {V^2\over\kappa^2}+\left({B_1\over \beta}
-\ti \nu_1\right){\Omega\over\kappa}={A_1\over\beta}
{V^2\over\kappa^2}+\left({B_1\over \beta} -(a_1+a_2 \cos^2({\bf
\hat{\Omega}}{\bf \hat{V}}))\right){\Omega\over\kappa}.%
 \end{equation}
\par
The stability analysis furnishes the same conclusions of the
parallel case, namely $L_2$ is stable for $V>V_{c2}$, $V_{c2}$ being
the second critical velocity defined by
\begin{equation}\label{VeloCrit2-gen}%
V_{c2}^2= {\beta\over A_1}
{\ti\nu_1^2-2\ti\nu_1\nu_2 \over \ti\nu_1-\nu_2} {\Omega \kappa}= {2 \ti\nu_1 \beta - B_1 \over A_1} {\Omega \kappa}%
\end{equation}
and depending on the angle between ${\bf \hat{\Omega}}$ and ${\bf
\hat{V}}$ through $\ti\nu_1$.  Looking at the first solution $L_1$
one notes that
\[
a_1\leq(a_1+a_2 \cos^2({\bf \hat{\Omega}}{\bf \hat{V}}))\leq
(a_1+a_2)=1
\]
which means
\[
{\tilde{\nu_1}}_{|V\perp \Omega}\leq {\ti \nu_1} \leq {\ti
\nu_1}^\parallel=\nu_1,
\] $\nu_1$ being the coefficient of the stationary solution $L_1$
in\;(\ref{solstazL1}) and plotted in Fig.\,\ref{confronto}. An
equivalent relation could also be written for $L_2$, but,
differently to $L_1$, coefficients $A_1$ and $B_1$ could depend on
the anisotropy of the tangle, namely  on the angle between $\bf V$
and $\bf \Omega$.

In Ref.\,\cite{Sciacca_PhisicaB_403}, the dimensionless quantities
appearing in the evolution equation\;(\ref{AltVinEqua}) are
determined by comparison with the experimental data in the steady
state, obtaining
\begin{equation}\label{ValueCosta}%
 {A_1\over\beta}=0.0125 , \hskip0.2in {B_1
\over\beta}= 3.90, \hskip0.2in \nu_1=2.036, \hskip0.2in \nu_2=0.97.%
\end{equation}
Instead, in  the case of pure counterflow $A_1/\beta=0.156$, which
confirms the dependence of the coefficients on the anisotropy of
the tangle \cite{Sciacca_PhisicaB_403}. By (\ref{ValueCosta}),
solutions (\ref{solstazL1}) and (\ref{solstazL2}) become
\be\label{solstazL1AValue}%
L_1^\parallel=2.036{\Omega\over\kappa}, \hspace{1cm} \textrm{and}
\hspace{1cm} L_2^\parallel= 0.0125
{V^2\over\kappa^2}+1.86{\Omega\over\kappa}.%
\ee These solutions are plotted in Fig.\,\ref{confronto}, where
the agrement with the experimental data is shown.

An interesting problem  is to establish which equation, either the
one based on the usual Vinen equation or the other one based on the
alternative Vinen equation, fits better the experimental data
obtained by Swanson, Barenghi and Donnelly \cite{SBD}. This
comparison was made in Ref.\,\cite{Sciacca_PhisicaB_403}: the two
stationary solutions in the range $V<V_{c2}$ represent the same
straight line in the plane $(L,V^2)$, whereas  in the range
$V^2>V^2_{c2}$ it is not the case. From the direct comparison with
the experimental data
--- by finding the errors between the stationary solutions of the
two models and the corresponding experimental values --- in
Ref.\;\cite{Sciacca_PhisicaB_403} the following statement  was
established: the stationary solution of the alternative Vinen
equation approaches better the experimental data (for
$V^2>V_{c2}^2$) than that of the usual Vinen equation.

In the same paper the unsteady behavior of the solutions of these
equations was also explored. Both equations exhibit remarkable
differences, such as the solutions of the alternative Vinen tend
much faster to their steady-state values, and this difference
depends on the value of the counterflow velocity. Even though
 detailed experimental data  on this unsteady behavior lacks, according to Swanson {\it et al.}
the time required to reach the steady state is less than 10 minutes
--- when the counterflow velocity $V$ is slightly above the critical
velocity $V_{c2}$ and it increases between two consecutive
experimental values  (see pag. 191, Ref.~\cite{SBD}). According to
the results of Ref.\;\cite{Sciacca_PhisicaB_403}, the temporal scale
of the solution of the usual Vinen equation was closer to the
observations than the temporal scale corresponding to the
alternative equation, which tends too fast to the final result.
Thus, it seems that the usual equation is preferable on these
grounds.

\section{Thermodynamic analysis}\label{Themodinamica}

In this Section  the general lines of Refs.~\cite{NF,JM-libro,JM2}
are followed to determine an evolution equation for ${\bf V}_s$
which is consistent with equation\;(\ref{AltVinEqua_gen}). This
situation is of special interest when ${\bf V}_s$ is not kept
constant, as it is usually assumed in (\ref{Vin}) or (\ref{Vin-al})
--- or in (\ref{VinEqua22}) and (\ref{AltVinEqua}). This more
general model may be useful to study the dynamic behavior of the
vortex tangle, when the counterflow velocity ${\bf V}$ varies in a
greater range than that considered in
Ref.\,\cite{Sciacca_PhisicaB_403}. According to the formalism of
nonequilibrium thermodynamics, evolution equations for ${\bf V}_s$
and $L$ can be obtained by writing ${d {\bf V}_s/ dt}$ and ${dL/
dt}$ in terms  of their conjugate forces $-\rho_s {\bf V}$ and
$\epsilon_V$. This is motivated also by the paper of Nemirowskii
{\it et al.} \cite{NF}, where for the entropy density $s$ of the
superfluid in the presence of vortex lines they considered the
differential form
\be\label{forma_diffe}%
T \frac{d s}{dt}  =-\rho_s{\bf  V}\cdot  \frac{d  {\bf V}_s}{d t}
+\epsilon_V \frac{dL}{dt},%
\ee where \be\label{cond_cond_coeff}%
-\rho_s{\bf  V}\equiv \frac{\partial u}{\partial {\bf V_s}},
\hskip0.4in \epsilon_V \equiv \frac{\partial u}{\partial L} =
\frac{\rho_s\kappa^2}{4\pi} \ln\left[\frac{1}{a_0
L^{1/2}}\right],%
\ee being $u$ the internal energy density and $\epsilon_V$ the
contribution to the internal energy per unit length of the vortex
line ($a_0$ is the dimension of the vortex core, which is very
small, of the order of one \AA) \cite{D1}.

First, recalling that in the presence of pure rotation (which
produces an ordered array of vortex  lines parallel to the
rotation axis) the evolution equation for ${\bf V}_s$ is
\cite{D1,JM2,HV}
\begin{equation}\label{equapervs} {d {\bf V}_s \over d t} + 2{\bf
\Omega} \times {\bf V}_s + {\bf i}_0 =-L {\rho_n\over\rho}
{B\over\tilde\alpha}
 \left( {\bf\widehat \Omega}  \times
{\bf\widehat \Omega} \times {\bf V} + {B'\over B} {\bf\widehat
\Omega} \times {\bf V} \right), \end{equation}
 where $\tilde\alpha=2/\kappa$,  ${\bf i}_0$ is  the inertial
force, $B$ and $B'$ are the Hall-Vinen dimensionless coefficients
describing the interaction between the normal fluid and the vortex
lines \cite{HV}.

In the presence of pure counterflow, the evolution equation for
${\bf V}_s$ can be written \cite{NF}
\begin{equation}\label{equasem} {d {\bf V}_s\over dt}={\rho_n
\over \rho}  A L {\bf V}, \end{equation}
 where  $A={2B/ 3 \tilde\alpha}$ \cite{D1}. Since in the steady state
 $L$ is proportional to $V^2$, the friction force on ${\bf V}_s$
 (which is opposite to ${\bf V}$), will be proportional to $V^3$, a
 result known as Gorter-Mellink law \cite{D1,BDV,NF}. Note, however, that this $V^3$ dependence is valid only
 at the steady state, whereas (\ref{equasem}) may be also applied in
 unsteady situations, in absence of rotation. Here we
 will be interested in the form of the
 friction force in unsteady situations including
 rotation and, in more general terms, in the general form of the
 evolution equation for ${\bf V}_s$.

In the simultaneous presence of counterflow and rotation, terms as
those in the right-hand side of (\ref{equapervs}) and
(\ref{equasem}) must be included in the evolution equation for
${\bf V}_s$, as well as an additional contribution ${\bf
F_{coupl}}={\bf F_c}({\bf V},{\bf \Omega})$ due to couplings
between counterflow, rotation and superfluid velocity. The
coupling term is found similarly to the Onsager's formalism of
non-equilibrium thermodynamics~\cite{NF,JM-libro,JM2}
\begin{equation}\label{equapervstensor} {d {\bf V}_s\over dt} +
2{\bf \Omega} \times {\bf V}_s + {\bf i}_0= {\rho_n\over\rho} L
{B\over\tilde\alpha} {\bf \Pi}\cdot {\bf V}  + {\bf F_c}({\bf
V},{\bf \Omega}), \end{equation}
 where   tensor
\begin{equation}\label{tensor}{\bf\Pi} \equiv (1-b) {2\over 3} {\bf U} +b
\left( {\bf U} - {\bf\widehat\Omega} {\bf\widehat\Omega}\right) +
b' {B'\over B} {\bf W}\cdot {\bf\widehat\Omega} \end{equation} was
introduced and studied in Ref.~\cite{JM4}.  In
Refs.~\cite{JM-libro,JM2}, for the sake of simplicity, $b=b'$ was
chosen. Here, $b$ and $b'$ are numerical parameters related to the
anisotropy and polarization of vortex tangle, describing the
relative weight of the array  of vortex lines parallel to
$\bf\Omega$ and the isotropic tangle:  $b=b'=0$ means an isotropic
tangle corresponding to pure counterflow and $b=b' =1$ means an
ordered array corresponding to pure rotation. The microscopic
physical interpretation of coefficients $b$ and $b'$ was made in
Ref.~\cite{JM4}. In particular $b$ is linked to the anisotropy of
the tangle and $b'$ to the polarity.

Our aim here is  to  include equations (\ref{AltVinEqua_gen}) and
(\ref{equapervstensor}) into a common thermodynamic framework, and
study possible couplings between them. But, because of the origin
of the term $\Omega V^2$ in equation (\ref{AltVinEqua}) is
unknown, as pointed out in Section\,\ref{sez311}, here special
emphasis is given to the two opposite cases: $a_1=1$ and $a_2=0$,
and $a_1=0$ and $a_2=1$. Of course, an intermediate situation
where both coefficients are non zero is not excluded and could be
an interesting physical situation which will not be considered for
the sake of simplicity. Future experiments could urge us to spend
further efforts on these intermediate cases.

\subsection{$\Omega V^2$ interpreted as $V {\bf \Omega}\cdot {\bf \hat{V}}{\bf
\hat{\Omega}} \cdot {\bf V}$ ($a_1=1$ and $a_2=0$)}\label{caso1}

In this subsection, term $\Omega V^2$ is assumed to come from $V
{\bf \Omega}\cdot {\bf \hat{V}}{\bf \hat{\Omega}} \cdot {\bf V}$
which depends only on the absolute values of the two fields. In
this spirit, equation\;(\ref{AltVinEqua_gen}) for $L$ with $a_1=1$
and $a_2=0$ is written in the second line of the following
system\;(\ref{sistema}). The second part of the evolution equation
for ${\bf V}_s$ is built up by means of the Onsager-Casimir
reciprocity relation, as in Refs.~\cite{JM-libro,JM2}. The result
is
\begin{equation}\label{sistema} \left[
\begin{matrix}
D {\bf V}_s\over dt\\
{d L\over dt}
\end{matrix}
\right]= \left[
\begin{matrix}
- {\rho_n \over \rho\rho_s}{B\over\tilde\alpha}L {\bf \Pi}  &
 {1\over\rho_s} {A_1\over\kappa}\left[ L-
\nu_1 {\Omega\over\kappa}\right]{\bf V} \\
-{1\over\rho_s} {A_1\over\kappa}\left[ L- \nu_1
{\Omega\over\kappa}\right]{\bf V} & - {1 \over \epsilon_V}\left[
\beta\kappa L^2 - B_1\left( L- \nu_2
{\Omega\over\kappa}\right)\Omega \right]
\end{matrix} \right] \left[\begin{matrix} -\rho_s{\bf V} \\ \epsilon_V
\end{matrix} \right],
 \end{equation}
where
\begin{equation}\label{equapervscompatta} {D {\bf V}_s\over dt}={d {\bf V}_s
\over d t} + 2{\bf \Omega} \times {\bf V}_s +{\bf i}_0.
\end{equation}  Therefore, in the presence of counterflow
and rotation the equation for $d{\bf V}_s/dt$ is
\begin{equation}\label{equapervsdefi}{d {\bf V}_s \over d t} +2
{\bf \Omega} \times {\bf V}_s +{\bf i}_0 = L{\rho_n\over\rho}
{B\over\tilde\alpha}
  {\bf \Pi }  \cdot {\bf V}
+{\epsilon_V\over\rho_s}
 {A_1\over\kappa}\left[ L-
\nu_1 {\Omega\over\kappa}\right]{\bf V}.
  \end{equation}
From a comparison with (\ref{equapervstensor}) the following
expression for the coupling friction force is obtained
\begin{equation}\label{fcoupl}%
{\bf F}_{coupl} = {\epsilon_V\over\rho_s}{A_1\over\kappa}\left[ L-
\nu_1
{\Omega\over\kappa}\right]{\bf V}. %
 \end{equation} This term is
proportional to $\bf V$, and is therefore a friction term (recall
that ${\bf V}$ is opposite to ${\bf V}_s$). Note the presence of a
negative term proportional to $\Omega$ and independent on $L$,
which means that the rotation of the vortex tangle reduces the
friction. This may be interpreted as a consequence of the
alignment of the vortices  to the rotation axis due to  ${\bf
\Omega}$, in such a way that they have less resistance to the flow
(recall that the friction is due to vortices orthogonal to the
counterflow velocity, but not to those parallel to it).

Substituting (\ref{solstazL1-gen}) and (\ref{solstazL2-gen}) in
the off-diagonal term in the matrix in (\ref{sistema}), one
obtains the following expression for the coupling force
\begin{equation}\label{fcouplmin}{\bf F}_{coupl}= {\epsilon_V\over\rho_s}
 {A_1\over\kappa}\left[ L_1-
\nu_1 {\Omega\over\kappa}\right]{\bf V} \equiv 0 , \hskip0.5in
\hskip0.5in\hbox{ for}~~V<V_{c2}, \end{equation}
\begin{equation}\label{fcouplmag}{\bf
F}_{coupl}={\epsilon_V\over\rho_s}
 {A_1\over\kappa}\left[ L_2-
\nu_1 {\Omega\over\kappa}\right]{\bf V}
 \equiv {\epsilon_V\over \rho_s} {A_1^2\over\beta \kappa^3}
 (V^2-V_{c2}^2) {\bf  V},  \hskip0.15in \hbox{ for}~~V>V_{c2}.
\end{equation} As a consequence, in a steady situation (and also in almost
steady state) for $V<V_{c2}$ the coupling force is absent (as in
pure rotation) while, for $V>V_{c2}$ --- when the array of
rectilinear vortex lines becomes a disordered tangle --- the
additional term (\ref{fcouplmag}) proportional to $\bf V$ appears.
Thus, $V_{c2}$ indicates the threshold not only of the vortex line
dynamics but also of the friction acting on the velocity ${\bf
V}_s$ itself.

\subsection{$\Omega V^2$ interpreted as $V {\bf \Omega}\cdot {\bf
\hat{\Omega}}{\bf \hat{V}} \cdot {\bf V}$ ($a_1=0$ and
$a_2=1$)}\label{caso2}

Here the same lines of the previous subsection are applied to the
opposite case $a_1=0$ and $a_2=1$, which means that term $\Omega
V^2$ comes from  $V {\bf \Omega}\cdot {\bf \hat{\Omega}}{\bf
\hat{V}} \cdot {\bf V}$, depending also on the angle between the two
vectors. So, writing again equation\;(\ref{AltVinEqua_gen}) with
$a_1=0$ and $a_2=1$ in the second line of the following
system\;(\ref{sistema1}), the second part of the evolution equation
for ${\bf V}_s$ is built up  by means of the Onsager-Casimir
reciprocity relation
\begin{equation}\label{sistema1} \left[
\begin{matrix}
D {\bf V}_s\over dt\\
{d L\over dt}
\end{matrix}
\right]= \left[
\begin{matrix}
- {\rho_n \over \rho\rho_s}{B\over\tilde\alpha}L {\bf \Pi}  &
 {1\over\rho_s} {A_1\over\kappa}\left[ L- \ti L_1\right]{\bf V}\\
-{1\over\rho_s} {A_1\over\kappa}\left[ L- \ti L_1\right]{\bf V} &
- {1 \over \epsilon_V}\left[ \beta\kappa L^2 - B_1\left( L- \nu_2
{\Omega\over\kappa}\right)\Omega \right]
\end{matrix} \right] \left[\begin{matrix} -\rho_s{\bf V} \\ \epsilon_V
\end{matrix} \right],
 \end{equation}
where $\ti L_1=\nu_1 {\Omega\over\kappa} \cos^2 ({\bf
\hat{\Omega}}{\bf \hat{V}})$.
 In this case, instead, the equation for $d{\bf V}_s/dt$ in the presence of counterflow
and rotation  is
\begin{equation}\label{equapervsdefi2}{d {\bf V}_s \over d t} +2
{\bf \Omega} \times {\bf V}_s +{\bf i}_0 = L{\rho_n\over\rho}
{B\over\tilde\alpha}
  {\bf \Pi }  \cdot {\bf V}
+{\epsilon_V\over\rho_s}
 {A_1\over\kappa}\left[ L- \nu_1 {\Omega\over\kappa}
\cos^2 ({\bf \hat{\Omega}}{\bf \hat{V}})\right]{\bf V}.
  \end{equation}
From a comparison with (\ref{equapervstensor}), the following
expression for the coupling friction force is obtained
\begin{equation}\label{fcoupl2}%
{\bf F}_{coupl} = {\epsilon_V\over\rho_s}{A_1\over\kappa}\left[ L-
\nu_1 {\Omega\over\kappa} \cos^2 ({\bf \hat{\Omega}}{\bf
\hat{V}})\right]{\bf V}.  %
\end{equation} Note that it is equal to (\ref{fcoupl}) unless $\cos^2 ({\bf
\hat{\Omega}}{\bf \hat{V}})$. So, all the observations below
relation\,(\ref{fcoupl}) are still valid. Note also that the
negative term, proportional to $\Omega$ and independent on $L$,
depends explicitly on the angle between $\bf V$ and ${\bf \Omega}$
and it is maximum for $\bf V$ parallel to ${\bf \Omega}$ whereas
it is zero for $\bf V$ orthogonal to ${\bf \Omega}$. The friction
term has always the same direction of counterflow velocity $\bf
V$, and in the  situation $\bf V\bot{\bf \Omega}$ depends only on
$V$ and $L$ but not on the angular velocity $\Omega$.

By using expressions  (\ref{solstazL1-gen}) and
(\ref{solstazL2-gen}),  the coupling term\;(\ref{fcoupl2}) in the
steady state becomes
\begin{equation}\label{fcouplmin_2}%
{\bf F}_{coupl}= {\epsilon_V\over\rho_s}
 {A_1\over\kappa}\left[ L_1-
\nu_1 {\Omega\over\kappa} \cos^2 ({\bf \hat{\Omega}}{\bf
\hat{V}})\right]{\bf V} \equiv 0 , \hskip0.5in \hskip0.5in\hbox{
for}~~V<V_{c2},%
 \end{equation}
\begin{equation}\label{fcouplmag_2}%
{\bf F}_{coupl}={\epsilon_V\over\rho_s}
 {A_1\over\kappa}\left[ L_2-
\nu_1 {\Omega\over\kappa} \cos^2 ({\bf \hat{\Omega}}{\bf
\hat{V}})\right]{\bf V}
 \equiv {\epsilon_V\over \rho_s} {A_1^2\over\beta \kappa^3}
 (V^2-V_{c2}^2) {\bf  V},  \hskip0.08in \hbox{ for}~~V>V_{c2}.%
\end{equation} Here, the conclusions are the same pointed out below
relations\;(\ref{fcouplmin}) and\;(\ref{fcouplmag}), and,
therefore,  also in this case $V_{c2}$ indicates the threshold of
the friction force acting on the velocity ${\bf V}_s$ itself.

\section{Conclusions}
The possibility of at least two reasonable evolution equations for
the vortex line density $L$ in superfluid turbulence was known
since the early days in which Vinen proposed them. This equation
was generalized to the combined situation of counterflow and
rotation, with $\bf V$ parallel to  $\bf \Omega$, into
(\ref{VinEqua22}) and (\ref{AltVinEqua}), which was compared with
the experimental data given by Swanson {\it et al.}.
In\;\cite{JM2} Vinen equation\,(\ref{VinEqua22}) was extended to
include the case $\bf V$ and  $\bf \Omega$ with different
directions. Here, following the same ideas of that paper, an
extension of the modified alternative Vinen
equation\,(\ref{AltVinEqua}) has been proposed
in\,(\ref{AltVinEqua_gen}). So, this equation  is valid when $\bf
V$ and $\bf \Omega$ have  indefinite directions, becoming the
evolution equation\,(\ref{AltVinEqua})  when these vectors are
parallel. The stationary solutions of this extended equation
differ from that of parallel case for the coefficient $\ti \nu$
instead of $\nu$. So, from the theoretical point of view one can
establish that an higher angle between $\bf \Omega$ and $\bf V$
implies a lower value for the steady state $L_1$, unless future
experiments will prove that $\ti \nu \equiv \nu$.

The case $\bf V$ orthogonal to $\bf \Omega$ was experimentally
carried out by Yarmchuck and Glaberson in Ref.\cite{YG}. They
arranged a pair of horizontal parallel glass plates  forming a
closed channel of rectangular cross-section closed at one end with
a heater nearby, and open at the other end to the liquid helium
bath. The channel was of large aspect ratio: the height being 0.5
mm, the width 1.4 cm, and the length 5.5 cm. The channel is
rotated around its vertical axis, orthogonal to the direction of
the heat flux, in such a way that the counterflow velocity $\bf V$
is orthogonal to angular velocity $\bf\Omega$. The authors check
temperature and chemical gradients as functions of heater power
and rotation speed. The results were linear regimes, in which the
temperature and chemical gradients increase when rotating velocity
is increased, and a critical heater power $q_{c2}$ (corresponding
a critical counterflow velocity $V_{c2}$), which increase as the
rotation speed increases, becoming proportional to $\sqrt\Omega$.

These results points out that coefficient $a_1$ in
equation\,(\ref{AltVinEqua_gen}) is not zero. For, in the situation
of the above experiment  $\bf V$ is orthogonal to $\bf\Omega$,
namely the term proportional to $a_2$ is zero because $\cos^2 (\bf
\hat{V} \bf \hat{\Omega})=0$. On the other side, experimental
results confirm the presence of the term proportional to
$V^2\Omega$. But, to establish whether the coefficient $a_2$ is
equal to zero, further experiments are required.

\begin{figure}
\centering \subfigure[]
{\includegraphics[width=5.5cm]{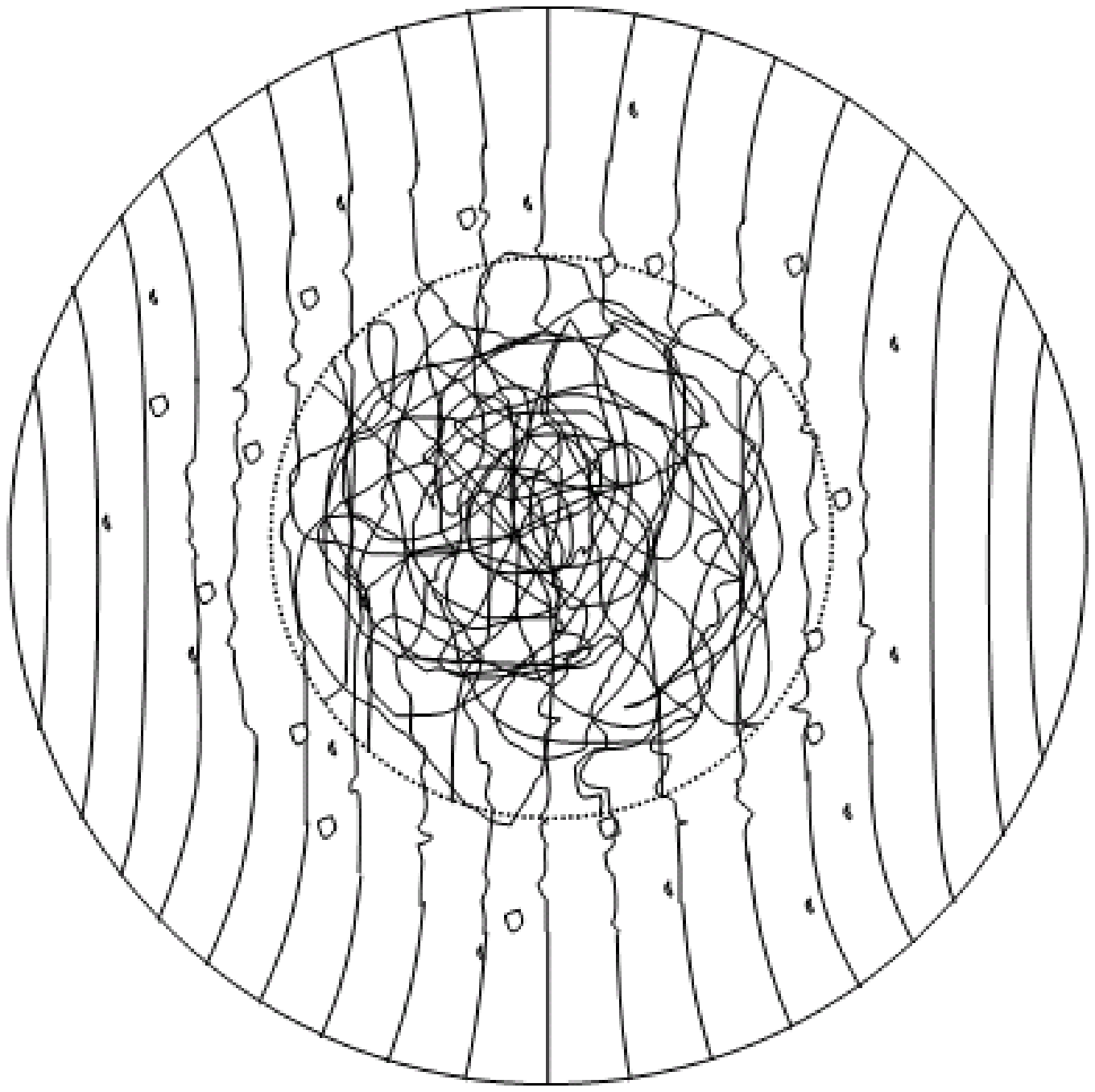}} \hspace{1mm}
\subfigure[] {\includegraphics[width=6cm]{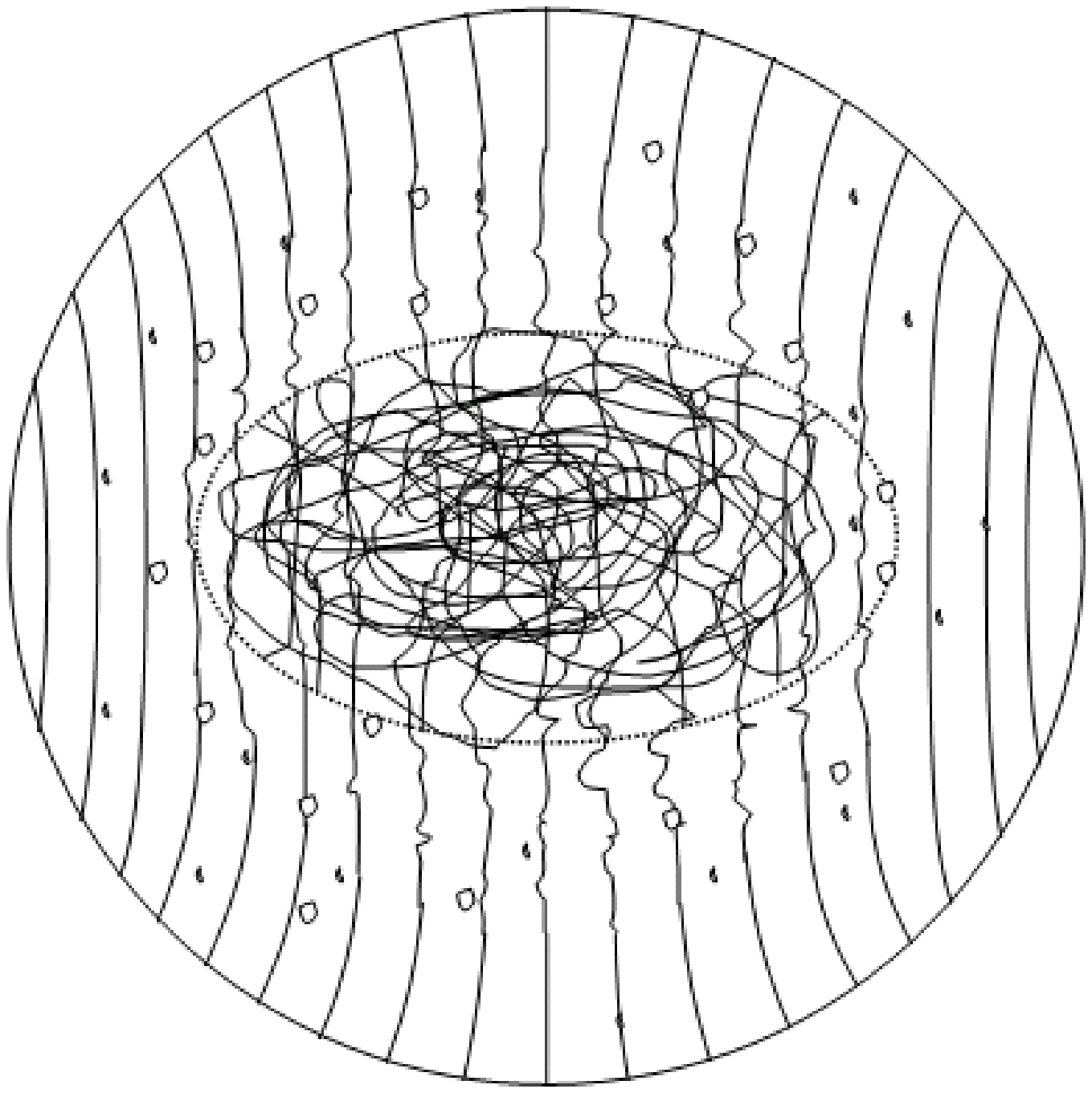}} \caption{
Sketch of a rotating sphere with an internal turbulent region and
an external laminar region. The boundaries of the turbulent
regions will be or not spherical depending on whether the critical
heat flux (\ref{VeloCrit2-gen}) does not depend on the relative
angle between heat flux and angular velocity or depends on it. The
picture of the external region ($r>r_2$) is only indicative, but a
strongly polarized turbulent region could be more
realistic.}\label{figura_stella}
\end{figure}

As mentioned in the Introduction of this paper, an interesting
application of these arguments could be a rotating spherical
container (like a star) filled by superfluid helium which suffers
the presence of radial heat flow (see figure\,\ref{figura_stella}).
Since  irradiation comes from the center of the star, the heat flow
will be proportional to $r^{-2}$, $r$ being the distance from the
center of the star. Because of the proportionality between heat flow
and counterflow velocity, this dependence on $r$ requires the
existence of a critical radius $r_2$ (corresponding to the second
critical velocity $V_{c2}$) which marks the turbulent region
($r<r_2$). Of course, the critical radius $r_2$ depends on the
angular velocity $\Omega$ from the relation (\ref{VeloCrit2-gen}).
An interesting question which arises could be whether $r_2$ depends
on the angle $\theta$ between ${\bf V}$ and ${\bf \Omega}$, or
similarly whether $a_2$ is zero or not. Because of the lack of
experimental evidence any conclusion on the presence of the
coefficient $a_2$ is not reached. However, the case $a_2=0$ would
require a critical radius free from $\theta$ (see relation
(\ref{VeloCrit2-gen})) and a spherical turbulent region of radius
$r_2$ as showed in figure\,\ref{figura_stella}a. The case $a_2\neq
0$, instead, would require a radius $r_2$ depending on $\theta$
caused by the fact that the critical counterflow velocity $V_{c2}$
increases when one moves from the equator to the poles of the
sphere. This means that $r_2(\theta)$ has a maximum at the equator
and minimum at the poles, that is the turbulent region is contained
in an ellipsoid (see figure\,\ref{figura_stella}b). Note
 that in both figures\,\ref{figura_stella} the region for $r>r_2$
 is filled of vortex array which suffers the perturbation coming
 from the inner turbulent region.

Because of the direct applications of these arguments on neutron
stars, some other remark could be pointed out. It is to note that
different forms of the superfluid turbulent core have implication
on the moment of inertia of the star which may influence the
acceleration periods. These acceleration periods are a real topic
which has deserved many interest for some researchers, but nothing
has still been  proved because the interior of the neutron stars
are not directly observable. But, some more information on neutron
stars can be obtained because Fig.\;\ref{figura_stella} can be
experimentally achieved: take a rotating sphere full of
superfluid, which has a small heated sphere in its centre, then
the details of the frontier between the vortex tangle and the
array of parallel vortices (or of a highly polarized tangle) can
be measured.

The procedure pointed out in Refs.~\cite{JM2} was generalized  to
derive an evolution equation for the superfluid velocity ${\bf
V}_s$, thermodynamically consistent with the proposed evolution
equation for $L$. The form (\ref{fcoupl}) and (\ref{fcoupl2}) point
out
--- in analogy with the conclusions reached in Ref.~\cite{JM2}  but in a more direct way
 --- that a tangle of a given $L$ and $\Omega$
has less resistance to the flow than a tangle with the same value of
$L$ but in absence of rotation. This  fact may be attributed to the
orientational influence of the rotation on the vortex lines.

In Ref.~\cite{JM2}, the authors obtained the coupling force ${\bf
F}_{coupl}$, between $\bf V$ and $\bf\Omega$, in the evolution
equation for superfluid velocity coupled to the generalized Vinen
equation. As pointed out along this paper, they distinguished
essentially two cases, both corresponding to ours. Regarding the
case of subsection\,\ref{caso2} --- in which the investigated term
of the generalized Vinen equation (and alternative Vinen equation)
depends on the angle between $\bf V$ and $\bf\Omega$ ---,
in~\cite{JM2} the authors find ${\bf F}_{coupl}\sim -L \cos({\bf
\hat{V}} {\bf \hat{\Omega}}) {\bf \Omega} /\sqrt{\Omega}$ which
differs from the corresponding (\ref{fcoupl2}) obtained in this
paper  for  the dependence on $V$. Note also that the sign of
coupling force\,(\ref{fcoupl2}) does not depend on the angle between
$\bf V$ and $\bf\Omega$ whereas that found by Jou and Mongiov\'{i}
does.  The coupling force\;(\ref{fcoupl2}) is, instead, similar to
the last two terms of equation\,(4.12) of paper~\cite{JM2}.

As regards the case of subsection\,\ref{caso1} --- in which the
investigated term does not depend on the angle between  $\bf V$
and $\bf \Omega$ --- it corresponds to case B of the
paper\,\cite{JM2}, where the authors found ${\bf F}_{coupl}\sim -L
\sqrt{\Omega} {\bf \hat{V}}$ which differs from that proposed
here. Again, the coupling force\,(\ref{fcoupl}) is similar to the
last two terms of equation\,(4.17) of paper~\cite{JM2}.

\subsection*{Acknowledgments} The author  acknowledges Professors
M.S. Mongiov\`{i} of the University of Palermo and D. Jou of the
University Autonoma de Barcelona for the highlighting discussions
on the arguments of the present paper.  Furthermore, the author
acknowledges the financial support from "Fondi 60\%" of the
University of Palermo and the "Assegno di ricerca" of the
University of Palermo.

\end{document}